\documentclass[prl,10pt,twocolumn,preprintnumbers,amsmath,superscriptaddress,amssymb,aps]{revtex4-1}

\usepackage{graphicx}% Include figure files
\usepackage{dcolumn}% Align table columns on decimal point
\usepackage{bm}% bold math
\usepackage{amsmath}
\usepackage{amssymb}
\usepackage{graphicx}
\usepackage{indentfirst}
\usepackage{booktabs}
\usepackage{multirow}
\usepackage{colortbl}
\usepackage{hyperref}% add hypertext capabilities
\usepackage{cleveref}
\usepackage[normalem]{ulem}
\selectfont
\begin{document}

\title{Disorder-Induced Spectral Splitting versus Rabi Splitting under Strong Light-Matter Coupling}
\author{Wei-Kuo Li}
\affiliation{Department of Chemistry and Biochemistry, University of Notre Dame, Indiana 46556, United States}

\author{Hsing-Ta Chen}
\email{hchen25@nd.edu}
\affiliation{Department of Chemistry and Biochemistry, University of Notre Dame, Indiana 46556, United States}

\date{\today}% It is always \today, today,

\begin{abstract}
%chatgpt, wkli
% Cavity assisted chemical dynamics is an emerging field at the intersection of quantum optics and chemical dynamics. Rabi splitting as a result of molecular transitions coupled to cavity mode is a hallmark for strong light-matter interaction within such systems. However, disorder effect in chemical systems are essential and largely understudied. In this study, we unveil a theoretical discovery wherein Rabi splitting manifests even in the regime of strong disorder. This phenomenon is elucidated through the coupling of an auxiliary effective dark mode to the cavity. Contrary to conventional assumptions, we explain that disorder not only does not suppress Rabi splitting in this regime but, intriguingly, enhances its manifestation. Furthermore, we establish that the scaling of Rabi splitting energy as $\sqrt{N}$, where N is the molecule count, persists within this disorder regime. We apply cavity QED (cQED) theory, Monte Carlo simulation, and Finite-Difference Time Domain (FDTD) simulation to estimate Rabi splitting in strong disorder regime for plasmonic-disk cavities, which agree well with current available experimental measurements.  Our findings suggest a paradigm shift for chemists, highlighting the importance of not solely focusing on highly-engineered, low-disorder systems but also considering the dynamics within "dirty" systems characterized by substantial disorder. This exploration opens new avenues for understanding and harnessing strong light-matter interactions in complex chemical environments.
The notion of strong light-matter coupling is typically associated with the observation of Rabi splitting, corresponding to the formation of the hybrid light-matter states known as polaritons. 
However, this relationship is derived based on the assumption that disorder can be ignored or acts as a perturbative effect.
Contrary to conventional treatment of disorder effects, we investigate the impact of strong disorder on the absorption spectrum by developing a non-perturbative effective model combined with classical electrodynamics simulation.
Intriguingly, we find that strong disorder leads to an enhanced spectral splitting that closely resembles Rabi splitting, yet originates from a fundamentally different mechanism as induced by the dark modes.
Specifically, we examine a disordered molecular ensemble in proximity to a plasmonic nanodisk and demonstrate disorder-induced spectral splitting in the absorption spectrum.
This conclusion raises a controversial issue, suggesting that both polaritons (dominate in the strong coupling regime) and dark modes (dominate in the strong disorder regime) can lead to spectral splitting, and one cannot distinguish them solely based on the steady-state absorption spectrum.
% This conclusion raises a controversial issue, indicating that strong disorder \wkle{alone?} can lead to spectral splitting and should not be ignored.
% This conclusion challenges the assumption that spectral splitting is a definitive hallmark of strong light-matter coupling, as it may instead be attributed to strong coupling disorder effects.
\end{abstract}

\maketitle

%\tableofcontents

%\section{\label{sec:level1}Introduction}
% \section{INTRODUCTION}
\paragraph{Introduction---}
A key focus of recent research in quantum light-matter interactions is on understanding molecular polaritons, which arise from the hybridization of photons absorption/emission with molecular excitations in the strong light-matter coupling regime\cite{ebbesen_hybrid_2016,ribeiro_polariton_2018,mandal_theoretical_2023,xiang_molecular_2024}.
Recent reports have demonstrated many potential applications of polariton in controlling chemical and material properties, such vibrational ground-state reactivity\cite{thomas_ground-state_2016,li_molecular_2022,du_catalysis_2022} and charge carrier mobility\cite{orgiu_conductivity_2015,xu_ultrafast_2023,balasubrahmaniyam_enhanced_2023,steger_long-range_2013,hou_ultralong-range_2020}, across various platforms such as Fabry-P\'erot microcavity and plasmonic nanostructures\cite{aberra_guebrou_coherent_2012,lidzey_photon-mediated_2000,thomas_tilting_2019,brawley_angle-independent_2021}.
It is widely accepted that the hallmark of strong light-matter coupling is Rabi splitting in the absorption spectrum, which emerges from the formation of upper/lower polariton states and is proportional to the effective coupling strength\cite{khitrova_vacuum_2006,george_ultra-strong_2015}.
However, a significant inconsistency remains between the observed spectral splitting and the underlying polariton-induced phenomena --- larger Rabi splitting does not always enhance the corresponding strong coupling effect\cite{chen_exploring_2024,fregoni_theoretical_2022,khazanov_embrace_2023}.

%disorder and dark states
Given this inconsistency, one might wonder: Is spectral splitting \emph{exclusively} a result of strong light-matter coupling? 
In the presence of weak disorder, the effective Rabi splitting can be either enhanced or suppressed, depending on the nature of the disorder (energetic or orientational, static or dynamical)\cite{gera_effects_2022,zhou_interplay_2023,climent_kubo-anderson_2024}.
It has been shown that disorder leads to unexpected phenomena, such as enhanced transport and entanglement\cite{allard_disorder-enhanced_2022,du_catalysis_2022,wellnitz_disorder_2022,tutunnikov_characterization_2024}.
To rationalize these phenomena, the prevalent explanation often involves the interaction between polaritons and dark states---molecular excitation states that remain uncoupled from the photon mode\cite{khazanov_embrace_2023,pandya_tuning_2022,botzung_dark_2020,upton_optically_2013}. 
Notably, most treatments of disorder consider the dark-state contribution as a perturbation that modulates the polariton states, thereby affecting the effective Rabi splitting.\cite{zhou_interplay_2023,perez-sanchez_simulating_2023}. 
Nevertheless, given that the photon mode interacts collectively with a large number of molecules and is subject to extensive disorder stemming from the spatial distribution of electromagnetic fields, molecular orientations, and environmental fluctuations, it seems improbable that perturbation theory can fully capture the impact of disorder on suppression/enhancement of spectral splitting.

In this letter, we reveal an unexpected spectral splitting as induced by strong disorder in light-matter coupling.
Remarkably, in spectroscopic measurements, this disorder-induced spectral splitting is qualitatively indistinguishable from Rabi splitting observed in an ordered system.
To elucidate this phenomenon, we develop non-perturbative approaches using collective bright and dark modes along with classical electrodynamics simulations.  
Our analytical solution predicts a phase diagram delineating the regimes of strong coupling and strong disorder where spectral splitting can be observed.
Furthermore, we consider a disordered molecular ensemble coupled to a plasmonic nanodisk and numerically demonstrate this disorder-induced effect.
\paragraph{Model---}
We consider an ensemble of $N$ molecules coupled to a plasmonic photon mode described by the total Hamiltonian
\begin{equation}
\begin{split} 
\hat{H} =& \hbar \Omega_p \hat{a}^\dagger_p \hat{a}_p + \sum_{j=1}^N \left( \hbar \Omega_v \hat{a}^\dagger_j \hat{a}_j + u_j^* \hat{a}^\dagger_p \hat{a}_j + u_j \hat{a}_p \hat{a}^\dagger_j \right)\\
&+{\cal E}_0 e^{-i\omega t} \Big( g_p \hat{a}^\dagger_p + \sum_{j=1}^N g_j \hat{a}^\dagger_j \Big)+c.c. 
\end{split}
\end{equation} 
Here the plasmonic cavity photon is characterized by the complex-valued frequency $\Omega_p=\omega_p-i{\gamma_p}/{\hbar}$ where $\gamma_p$, $\omega_p$ and $\hat{a}_p$ are the line width, frequency and annihilation operator for the cavity photon mode.
The molecule ensemble comprises a set of identical molecules with $\Omega_v=\omega_v-i{\gamma_v}/{\hbar}$ where $\hbar\omega_v$ is the molecular excitation energy and $\gamma_v$ is the relaxation rate. 
We denote $\hat{a}_j$ as the annihilation operator of the $j$-th molecule and neglect the inter-molecular interactions and dipole self-energy terms\cite{flick_atoms_2017,mandal_polariton-mediated_2020}.
The molecule-plasmon coupling ($u_j$) explicitly depends on $j$ due to the spatial distribution of the electromagnetic fields and molecular orientations beyond the long-wavelength approximation.
%Driving field
Both the molecular systems and plasmonic cavity mode are driven by an incident field with the driving amplitude ${\cal E}_0$ and the driving frequency $\omega$.
The plasmonic photon mode is coupled to the incident field by $g_p$ and the molecule-field coupling is $g_j$.
Note that we employ the rotating wave approximation.

% \subsection{Equation of Motion and Absorption Spectrum}
With the model Hamiltonian, we can derive the equation of motion for their expectation values, $\langle \hat{a}_p\rangle\equiv a_p(t)$ and $\langle \hat{a}_j\rangle\equiv a_j(t)$, using the Heisenberg equation and Ehrenfest theorem. 
We use the steady state anstaz $a_p(t)=\bar{a}_p e^{-i\omega t}$ and $a_j(t)=\bar{a}_j e^{-i\omega t}$ and the steady-state amplitudes $\bar{a}_p$ and $\bar{a}_j$ satisfy the following equation of motion (EOM)
\begin{align}
i(\Omega_{p}-\omega)\bar{a}_p + \frac{i}{\hbar}\sum_{j=1}^{N}u_{j}^*\bar{a}_j &=	-\frac{i}{\hbar}g_{p}{\cal E}_0 \label{eq:CQED-EOM-1}\\
i(\Omega_{v}-\omega)\bar{a}_j + \frac{i}{\hbar}u_{j}\bar{a}_p &=	-\frac{i}{\hbar}g_{j}{\cal E}_0 \label{eq:CQED-EOM-2}
\end{align}
The absorption spectrum can be computed in terms of the steady-state amplitude by 
\begin{equation}\label{eq:CQED-Pabs}
P(\omega)=-2\omega\text{Im}\Big[{\cal E}_0^*\Big(g_p^* \bar{a}_p  +\sum_j g_{j}^*  \bar{a}_j\Big)\Big]
\end{equation}
Note that the steady-state absorption spectrum is derived using the energy loss rate of the incident field \cite{SM}.%\todo{(see SM XXX)}

% \subsection{Effective Model Technique} 
\paragraph{Equations of motion for collective modes---}
In the presence of disorder, we express the molecule-field and molecule-plasmon coupling in the form of $g_{j}=g_0+\delta{g}_{j}$ and $u_{j}=u_0+\delta{u}_{j}$.
The average coupling strength is given by $g_0=\frac{1}{N}\sum_j g_{j}$ and $u_0=\frac{1}{N}\sum_j u_{j}$. %i.e. $\sum_j\delta{g}_{j}=\sum_j\delta{u}_{j}=0$. 
The coupling disorder is characterized by their variance $\sigma_{g}^{2}=\frac{1}{N}\sum_{j=1}^{N}|\delta{g}_{j}|^{2}$ and $\sigma_{u}^{2}=\frac{1}{N}\sum_{j=1}^{N}|\delta{u}_{j}|^{2}$, respectively.
Notably, the correlation between $\delta{g}_{j}$ and $\delta{u}_{j}$ is quantized by the co-variance $\langle\delta{g}^*\delta{u}\rangle=\langle\delta{u}^*\delta{g}\rangle^*=\frac{1}{N}\sum_{j=1}^{N}\delta{g}_{j}^*\delta{u}_{j}\equiv \xi\sigma_u\sigma_g$ where the correlation parameter $\xi$ is a complex number with $|\xi|\le1$.

From Eqs.~\eqref{eq:CQED-EOM-1} and \eqref{eq:CQED-EOM-2}, we define the following collective modes: 
(\emph{i}) the bright mode $\bar{B}=\frac{1}{\sqrt{N}}\sum_{j}\bar{a}_{j}$, 
(\emph{ii}) the dark mode for the molecule-plasmon coupling
$\bar{D}_u=\frac{1}{\sigma_{u}\sqrt{N}}\sum_{j=1}^{N}\delta{u}_{j}^*\bar{a}_{j}$, and (\emph{iii}) the dark mode for the molecule-field coupling
$\bar{D}_g=\frac{1}{\sigma_{g}\sqrt{N}}\sum_{j=1}^{N}\delta{g}_{j}^*\bar{a}_{j}$.
These collective modes satisfy the effective EOM
\begin{align}
    &\begin{pmatrix}
    \hbar(\omega-\Omega_p) & \sqrt{N}u_0^* & \sqrt{N}\sigma_{u} & 0 \\
    \sqrt{N}{u_0} & \hbar(\omega-\Omega_v) & 0 & 0 \\
    \sqrt{N}\sigma_{u} &  0 & \hbar(\omega-\Omega_v) & 0 \\
    \sqrt{N}\sigma_{u}\xi & 0 & 0 & \hbar(\omega-\Omega_v)
    \end{pmatrix}
    \begin{pmatrix}
    \bar{a}_p \\ \bar{B} \\ \bar{D}_u \\ \bar{D}_g
    \end{pmatrix} \nonumber \\
&=   {\cal E}_0\begin{pmatrix}
    g_p \\ \sqrt{N}g_0\\ \sqrt{N}\sigma_{g}\xi^*\\ \sqrt{N}\sigma_{g}
    \end{pmatrix}\label{eq:effective-EOM}
\end{align}
% \begin{equation}\label{eq:effective-EOM}
%     (\hbar\omega \mathbf{I}- \mathbf{H})\mathbf{A}={\cal E}_0\mathbf{F}
% \end{equation} where 
% \begin{equation}\label{eq:effective-Hamiltonian}
%     \mathbf{H}= 
%     \begin{pmatrix}
%     \hbar\Omega_p & \sqrt{N}u_0^* & \sqrt{N}\sigma_{u} & 0 \\
%     \sqrt{N}{u_0} & \hbar\Omega_v & 0 & 0 \\
%     \sqrt{N}\sigma_{u} &  0 & \hbar\Omega_v & 0 \\
%     \sqrt{N}\sigma_{u}\xi & 0 & 0 & \hbar\Omega_v
%     \end{pmatrix}
% \end{equation}
% and $\mathbf{F}=(g_p, \sqrt{N}g_0, \sqrt{N}\sigma_{g}\xi^*, \sqrt{N}\sigma_{g})^T$, $\mathbf{A}=(\bar{a}_p, \bar{B}, \bar{D}_u, \bar{D}_g)^T$.
and the absorption power can be expressed as 
\begin{equation}\label{eq:effective-Pabs}
P(\omega)= -2\omega\text{Im}\left [  {\cal E}_0^* \left ( g_p^*\bar{a}_p + \sqrt{N}g_0^*\bar{B} + \sqrt{N}\sigma_g\bar{D}_g \right ) \right ]. 
\end{equation}
Note that, for given values of $\{g_j\}$ and $\{u_j\}$, solving Eq.~\eqref{eq:effective-EOM} is equivalent to solving Eqs.~\eqref{eq:CQED-EOM-1} and \eqref{eq:CQED-EOM-2}.
We emphasize that the absorption spectrum is expressed in terms of $\bar{a}_p$, $\bar{B}$ and $\bar{D}_g$.

Under the resonant condition ($\omega_p=\omega_v$), the key insight we gain from the collective-mode EOM is as follows. 
We notice that the $4\times4$ matrix in Eq.~\eqref{eq:effective-EOM} has two degenerate eigenvalues $\hbar\Omega_v$ and two separated eigenvalues $\hbar\Omega_{\pm}=\hbar\Omega_v\pm\sqrt{N(|u_0|^2+\sigma_u^2)}$.
The analytical expression of $P(\omega)/\omega$ is a combination of a Lorentzian peak centered at $\hbar\Omega_v$ and two side peaks centered at $\hbar\Omega_\pm$\cite{SM}.%\todo{(see SM. XXX)}
Importantly, the energy separation between the side peaks is $\sqrt{N(|u_0|^2+\sigma_u^2)}$, indicating that the spectral splitting depends on both the bright-mode coupling strength ($|u_0|$) and the dark-mode coupling strength ($\sigma_u$). 
We emphasize that observing spectral splitting alone is insufficient to determine whether the splitting arises from $|u_0|$ or $\sigma_u$.
Additionally, we note that the $\sqrt{N}$ scaling of the energy separation persists regardless of whether the splitting is due to $|u_0|$ or $\sigma_u$.
\paragraph{Electrodynamics simulation---}
In a more realistic system, the coupling parameters $g_p$, $\{g_j\}$, $\{u_j\}$ can be estimated by employing classical electrodynamics simulations\cite{oskooi_meep_2010}.
Firstly, we numerically calculate the absorption spectrum in the absence of the molecules ($P_\text{num}(\omega)$)\cite{SM}.%\todo{(see SM~XXX)}.
Then we can estimate $g_p$, $\omega_p$, and $\gamma_p$ by fitting the absorption spectrum to the functional form (derived from Eq.~\eqref{eq:CQED-EOM-1} with $u_j=0$)
\begin{equation}
     P_\text{num}(\omega) =  \frac{2\omega\gamma_p|g_p {\cal E}_0|^2}{[\hbar (\omega - \omega_p )]^2 + \gamma_p^2}  
\end{equation}
Meanwhile, the classical electrodynamics simulation provides the total field $\bm{\mathcal{E}}_\text{tot}$ and the incident electric field $\bm{\mathcal{E}}_\text{in}$. 
We approximate the local plasmonic field as $\bm{\mathcal{E}}_\text{loc}=\bm{\mathcal{E}}_\text{tot}-\bm{\mathcal{E}}_\text{in}$, assuming that the far-field scattering is weakly coupled to molecules near the plasmonic nanostructure. 
In addition, we neglect the radiative feedback from molecular emissions, so the local field is generated solely by the plasmon mode, i.e. $\bm{\mathcal{E}}_\text{loc} \propto \bar{a}_p$  \cite{miwa_quantum_2021}.

Next, to calculate the absorption spectrum in the presence of a molecular ensemble, we introduce three approaches.

({I}) \emph{Homogeneous and isotropic approximation}: %FDTD-INT[Effective Model]
We assume the molecular distribution is homogeneous and the molecular orientation is isotropic and uncorrelated with the position.
Thus, we can treat the molecular ensemble as a continuum with the average coupling strength $g_0=u_0=0$ (due to the isotropic assumption) and numerically calculate $\sigma_g^2 = {|\mu|^2}/{3}$ and \cite{SM}%\todo{(see SM. XXX)}
\begin{align}
    % \sigma_u^2&=\frac{|\mu|^2 }{3|\bar{a}_p|^2} \langle|\bm{\mathcal{E}}_\text{loc}|^2 \rangle_\mathbf{r} \\
    \sigma_u^2&=\frac{ \gamma_p^2|\mu|^2}{3|g_p|^2}\frac{\langle |\bm{\mathcal{E}}_\text{loc}|^2 \rangle_\mathbf{r}}{|{\cal E}_0|^2 }   \label{eq:int-u} \\ 
    \xi &= \frac{ \langle \bm{\mathcal{E}}_\text{loc}^* \cdot \bm{\mathcal{E}}_\text{in}  \rangle_\mathbf{r}}{\sqrt{\langle |\bm{\mathcal{E}}_\text{loc}|^2 \rangle_\mathbf{r}}{\sqrt{\langle |\bm{\mathcal{E}}_\text{in}|^2 \rangle_{\mathbf{r}}}} } \label{eq:int-xi}
\end{align}
Here $\mu$ is the amplitude of the transition dipole moment and $\langle f(\mathbf{r})\rangle_\mathbf{r}=\frac{1}{|\mathcal{V}|}\int_\mathcal{V} d\mathbf{r}^3f(\mathbf{r})$ denote the average within a chosen volume $\mathcal{V}$. 
With $\sigma_g$, $\sigma_u$, and $\xi$ obtained by numerical integration of the electric fields, we can calculate $P(\omega)$ by solving Eqs.~(\ref{eq:effective-EOM}--\ref{eq:effective-Pabs}).

({II}) \emph{Monte Carlo Method}: % FDTD-MC
We randomly sample molecular position $\mathbf{r}_j$ and dipole orientation $\mathbf{n}_j$ in close proximity to a plasmonic nanostructure to generate $\{g_j\}$ and $\{u_j\}$ by
\begin{align}
    g_{j} &=  \frac{\bm{\mathcal{E}}_\text{in}(\mathbf{r}_j,\omega_p)}{{\cal E}_0}\cdot\mu^*\mathbf{n}_j \label{eq:g_j_FDTD}\\
    % u_{j} &= \frac{\bm{\mathcal{E}}_\text{loc}(\mathbf{r}_j,\omega_p) }{a_p} \cdot \mu^* \mathbf{n}_j \\
    %u_{j} &=\frac{\bm{\mathcal{E}}_\text{loc}(\mathbf{r}_j,\omega_p) }{\bar{a}_p} \cdot \mu^* \mathbf{n}_j  \label{eq:u_j_FDTD}
    u_{j} &=i\frac{\gamma_p}{g_p}\frac{\bm{\mathcal{E}}_\text{loc}(\mathbf{r}_j,\omega_p) }{{\cal E}_0} \cdot \mu^* \mathbf{n}_j  \label{eq:u_j_FDTD}
\end{align}
Note that $\bm{\mathcal{E}}_\text{in}(\mathbf{r}_j,\omega_p)$ and $\bm{\mathcal{E}}_\text{loc}(\mathbf{r}_j,\omega_p)$ indicate their Fourier component at frequency $\omega_p$ respectively.
With $\{g_j\}$ and $\{u_j\}$ for molecular ensemble, we can solve Eqs.~(\ref{eq:CQED-EOM-1}--\ref{eq:CQED-EOM-2}) by direct matrix inversion and calculate the absorption spectrum by  Eq.~\eqref{eq:CQED-Pabs}.

({III}) \emph{Collective-mode EOM}: % FDTD-RG
With the sampled $\{g_j\}$ and $\{u_j\}$, we can compute $g_0$, $\sigma_g$, $u_0$, $\sigma_u$, $\xi$.
Thus, we can solve the effective EOM Eqs.~(\ref{eq:effective-EOM}) and calculate the absorption spectrum by Eq.~\eqref{eq:effective-Pabs}.
As a final note, this approach should converge to the Monte Carlo result in the large $N$ limit.

% \section{RESULTS AND DISCUSSION}
% \subsection{Statistical Properties of a Toy Model: Randomly Oriented Ensemble }
\paragraph{Disorder-Induced Spectral Splitting---}
Before calculating the nanodisk spectrum, it is useful to gain insights into the effective coupling model.
For simplicity, we choose $g_p=0.03$ as a real-valued parameter (subject to a unitary transformation)\cite{SM}.
First, we consider a closed cavity ($g_j=0$, i.e. $g_0=0$ and $\sigma_g=0$) where the external field is not directly coupled to the molecules.
We assume that the molecule-plasmon coupling takes the form $u_j=W_j\cos(\theta_j)$.
Here $W_j$ is a random variable following a Gaussian distribution of mean $\bar{W}$ and width $\Delta{W}$ for modeling the spatial variation of the electromagnetic field; $\theta_j$ is a uniform random variable in $[-\theta_\text{max},\theta_\text{max}]$ for modeling the molecular orientation. 
For the resonance condition, we use $\gamma_p=\gamma_v=1$ as a unit of energy and  $\hbar\omega_p/\gamma_p=\hbar\omega_v/\gamma_v=10$. 
We choose $\bar{W}=1.2$ and consider the long-wavelength approximation (LWA) case ($\Delta{W}=0$) and the non-uniform field case ($\Delta{W}=1.1$).

\begin{figure} % Optional placement specifier
    \centering
    \includegraphics[width=1.05\linewidth]{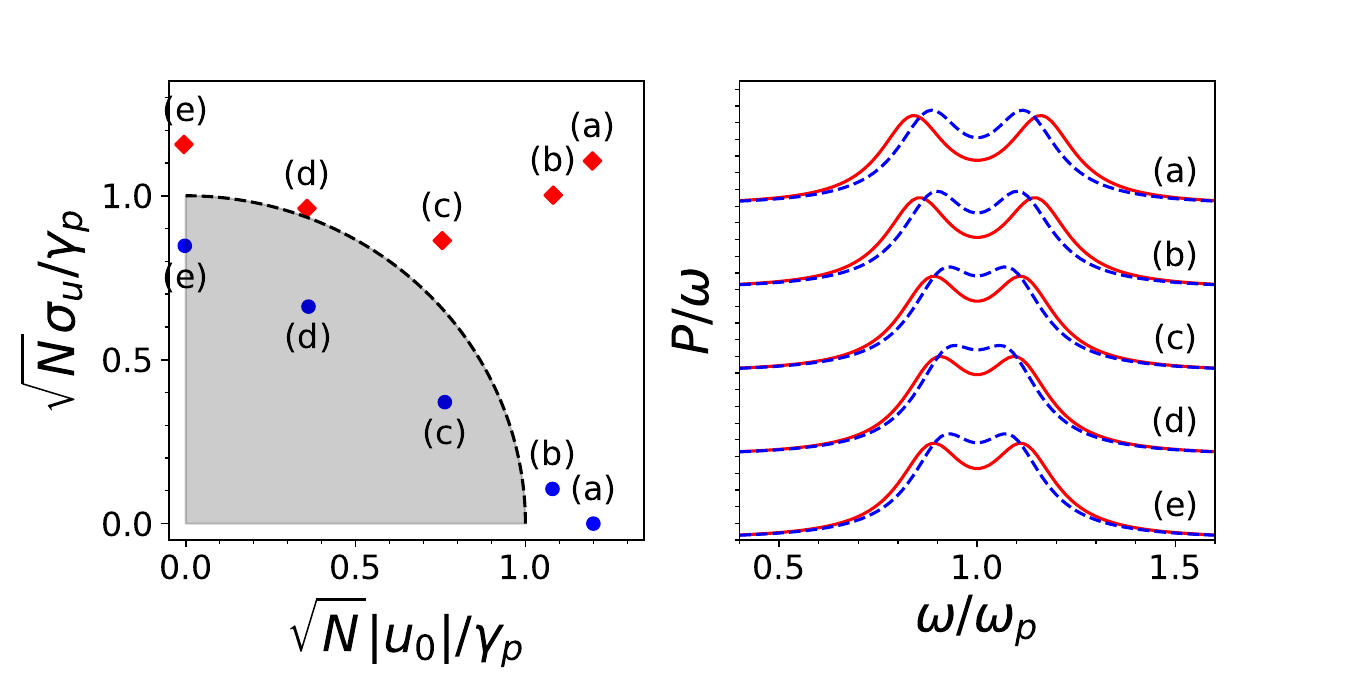} % Adjust width as needed
    \caption{Left: The phase diagram in terms of $\sqrt{N}|u_0|/\gamma_p$ and $\sqrt{N}\sigma_u/\gamma_p$. The dashed line depicts the criterion $N(|u_0|^2+\sigma_u^2)=\gamma_p^2$ and the shaded area indicates the parameter regime where the two peaks cannot be resolved. 
    % The spectral splitting is caused by strong \htc{collective} coupling when $|u_0|>\sigma_u$, and by strong \htc{disordered} coupling when $|u_0|<\sigma_u$.
    The orientation disorder is swept through $\theta_\text{max}= 0, \frac{1}{4}\pi, \frac{1}{2}\pi, \frac{3}{4}\pi, \pi$, labeled by (a)--(e) for the LWA cases ($\Delta{W}=0$, blue circles) and the non-uniform field cases ($\Delta{W}=1.1$, red diamonds). 
    Right: The corresponding absorption spectra $P(\omega)/\omega$ are plotted for the LWA (blue dashed lines) and the non-uniform field (red solid lines) cases.
    Note that, while (e) indicates completely disordered photon-molecule coupling (i.e. $|u_0|=0$), we can resolve two absorption peaks when the coupling disorder is large.
    More importantly, although blue (a) and red (e) arise from different mechanisms, they exhibit nearly identical absorption spectra.}
    \label{fig:fig1} % Optional label for cross-referencing
\end{figure}

Fig~\ref{fig:fig1} illustrates a phase diagram in terms of $u_0$ and $\sigma_u$ and their corresponding absorption spectra.
For completely aligned molecules ((a), $\theta_\text{max}=0$), both cases exhibit a spectral splitting larger than $\gamma_p$, but caused by different mechanisms.
In the LWA case, the energy separation is attributed to strong bright-mode coupling ($\sqrt{N}|u_0|>\gamma_p$), indicating that the spectral splitting is due to strong light-matter coupling (i.e. Rabi splitting).
However, in the non-uniform field case, the energy separation stems from both the bright-mode coupling and the dark-mode coupling ($\sqrt{N(|u_0|^2+\sigma_u^2)}>\gamma_p$).
Notably, for $\theta_\text{max}\ge\frac{1}{2}\pi$ (c)--(e), the LWA cases are within the regime $\Omega_R/\gamma_p<1$ and the two peaks cannot be resolved (shown in the right panel).

For molecules with large orientational disorder ((e), $\theta_\text{max}=\pi$), the bright-mode coupling strength becomes zero (i.e. $|u_0|=0$), and the absorption spectrum solely results from the dark-mode coupling $\sigma_u$.
In the right panel of Fig~\ref{fig:fig1}, we observe that the disorder-induced spectral splitting in the corresponding spectrum for the non-uniform field case (red~(e)). 
More importantly, we find that the red~(e) spectrum is almost identical to the blue~(a) spectrum (obtained in the LWA due to strong bright-mode coupling).
This result suggests that the observed spectral splitting alone is insufficient to reliably identify strong light-mater coupling.

% The results of the effective models are plotted in Fig. \ref{fig:fig2} where two Cases are discussed: {\bf Case I} field-disorder free case(blue dots): $g_0=1.3$, $\sigma_g=0$, {\bf Case II} field disordered case(red diamonds):  $g_0=1.3$, $\sigma_g=1.1$. In both cases the orientation disorder is swept through $\theta_{max} = 0,\frac{1}{4}\pi,\frac{1}{2}\pi,\frac{3}{4}\pi, \pi$. In Case I the increasing orientation disorder caused the system to crossover from the strong coupling regime to the weak coupling regime(as shown in Fig. \ref{fig:fig1-1}(a)), while the two-peaks profile of Rabi splitting in absorption spectra is merged into one(as shown in Fig. \ref{fig:fig1-1}(b) blue dashed line from top to bottom). On the other hand, Case II also show that Rabi splitting is suppressed (Fig. \ref{fig:fig1-1}(b) red solid line) but the system remains in strong coupling regime(Fig. \ref{fig:fig1-1}(a)). This suggests that in this model while orientation-disorder suppresses Rabi splitting, field-disorder can increase Rabi splitting. This supports the idea of Disorder driven Rabi Splitting, as an increment in field-disorder not only does not kill the Rabi splitting but enhanced it. We also want to point out that the Rabi splitting can survive at $\bar{g_c}=0$, where the system is maximum orientation-disordered.

\begin{figure} % Optional placement specifier
    \centering
    \includegraphics[width=1.05\linewidth]{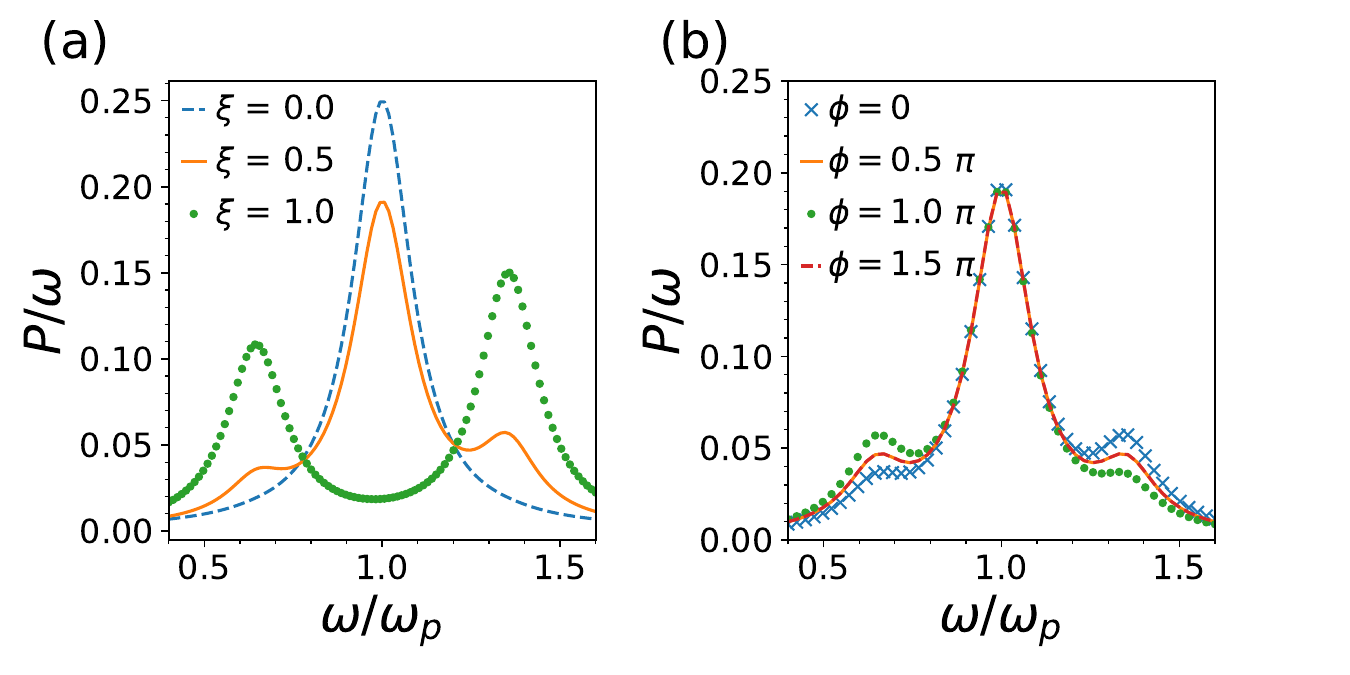} % Adjust width as needed
    \caption{
    The absorption spectrum ($P(\omega)/\omega$) of a disordered molecular ensemble as obtained by solving the collective mode EOM. We choose $\sqrt{N}\sigma_u \gg g_p$ and fix $\sqrt{N}\sigma_u$ and $\sqrt{N}\sigma_g$ and vary the correlation parameter $\xi=|\xi|e^{i\phi}$.  %, $\Omega_p=\Omega_v=10-i$, and $N=10^5$.
    (a) For $\phi=0$, $P(\omega)/\omega$ exhibit a single peak at $\omega=\omega_p$ for the uncorrelated case ($|\xi|=0$) and two peaks at $\omega=\Omega_\pm$ for the fully correlated case ($|\xi|=1$).
    (b) For $|\xi|=0.5$, the spectra with varying relative phase $\phi=0,0.5\pi,\pi,1.5\pi$ shows that the as induced by the interference between $a_p$ and $D_g$ modes. 
    % \htc{The effect of correlated field and cavity coupling constants in classical effective models. The coupling-constants correlator $\left \langle  {\delta g_c}^*\delta g_v  \right \rangle = \sigma_{gc}\sigma_{gv} \left |r\right | exp [ i \phi ]$. (a) $r$ is 0 (uncorrelated), 50\%  (partially correlated), and 100\% (fully correlated) while the phase $\phi$ is set to 0. (b) $r$ is 50\%  (partially correlated), and phase $\phi$ is 0, $0.5\pi$, $\pi$, $1.5\pi$.}
    }
    \label{fig:fig2} % Optional label for cross-referencing
\end{figure}

To further analyze the disorder-induced spectral splitting, we turn our attention to an open cavity with a completely disordered molecular ensemble.
We set $g_0=u_0=0$ and choose the coupling variance to be $\sqrt{N}\sigma_u/\gamma_p=3.5$ and $\sigma_g/\sigma_u=0.1$, and vary the correlation parameter $\xi=|\xi|e^{i\phi}$. 
% The plasmon-field coupling is $g_p=0.03$ and the molecule number is $N=10^5$.
In Fig. \ref{fig:fig2}(a), we differentiate the spectra for the uncorrelated($|\xi|=0$), partially correlated ($|\xi|=0.5$), and fully correlated coupling ($|\xi|=1$).
For the uncorrelated case ($|\xi|=0$), since the $D_u$ and $D_g$ modes are orthogonal to each other, the central peak is contributed by the $D_g$ mode (proportional to $\sqrt{N}\sigma_g$) and the side peaks are contributed only by the plasmonic mode $a_p$ (scaled with $g_p$). 
Thus, the central peak dominates as we choose $\sqrt{N}\sigma_g\gg g_p$.
For a partially correlated case ($|\xi|=0.5$), three peaks are visible as a result of mixing the $a_p$ and $D_g$ modes.
For the fully correlated coupling ($|\xi|=1$), the central peak vanishes and the side peaks are dominated by the $D_g$ mode, manifesting disorder-induced spectral splitting.
In Fig. \ref{fig:fig2}(b), we show that the interference of the side peak as the relative phase of the correlation parameter varies.  
Since we chose $g_p$ to be a real value, this interference between the $a_p$ and $D_g$ modes is controlled by the relative phase $\phi$.
% \htc{We need to justify $1/\sqrt{N}$ by choosing parameters.}

% In this case, the bright ($\bar{B}$) mode does not contribute to the absorption spectrum and we can derive the analytical expression: 
% \begin{equation}
%     % P(\omega)=-|{\cal E}_0|^2\omega\text{Im}\left[\frac{2N\sigma_g^2(1-|\xi|^2)}{\omega-\Omega_p}+\sum_{\lambda=\pm}\frac{|g_p+\lambda\sqrt{N}\sigma_g\xi^*|^2}{\omega-\Omega_\lambda}\right]
%     \frac{P(\omega)}{\omega}\propto\text{Im}\left[\frac{2N\sigma_g^2(1-|\xi|^2)}{\omega-\omega_p-i\gamma_p/\hbar}+\sum_{\lambda=\pm}\frac{|g_p+\lambda\sqrt{N}\sigma_g\xi^*|^2}{\omega-\Omega_\lambda-i\gamma_p/\hbar}\right]
% \end{equation}
% where $\Omega_\pm=\omega_p\pm\sqrt{N(|u_0|^2+\sigma_u^2)}$.

% \subsection{FDTD Simulations of Circular Metal Disk as Plasmonic cQED System}
\paragraph{Molecular ensemble coupled to a plasmonic nanodisk---}
With the insight we gained from the effective coupling model, we are now ready to analyze the absorption spectrum as we obtained by classical electrodynamics simulations.
Here we consider a gold disk of radius $0.25\ \mu\text{m}$ and thickness $0.0125\ \mu\text{m}$ placed on the $y$-$z$ plane.
% Incident field
The incident field is chosen to be linearly polarized in the $z$ direction and propagating toward the $+x$ direction.
We employ the finite-difference time-domain (FDTD) method to calculate the absorption spectrum without molecular ensemble using a short pulse and $\mathcal{E}_0=1 [V/m]$\cite{SM,taflove_computational_2005,oskooi_meep_2010,sukharev_efficient_2023}. %\todo{(see SM. XXX)}
% To calculate the absorption spectrum, we assume a short pulse of the form $A\cos(\omega_p t)e^{-(bt)^2}$ where $A$ is chosen by normalizing the absorption spectrum so that  $\mathcal{E}_0=1 [V/m]$ and $b$ is sufficiently large so that the field strength are approximately a constant near $\omega_p$.\cite{nitzan_chemical_2006} 
% \wkl{should we use $Ae^{-(bt)^2-i\omega_p t}$. I did not normalize A according to the sentence above, also large b is not tested put it supplement material cite MEEP here and say we cal. $P(w)$ @ ${\cal E}_0$=1 [V/m]}
% We employ the finite-difference time-domain (FDTD) method to calculate the electromagnetic fields without the nanodisk ($\bm{\mathcal{E}}_\text{in}$) and in the presence of the nanodisk ($\bm{\mathcal{E}}_\text{tot}$).\cite{oskooi_meep_2010} 
By fitting the absorption spectrum in the absence of the molecular ensemble, we can estimate the plasmonic frequency $\hbar\omega_p\approx6196\ \text{cm}^{-1}$ and the linewidth $\gamma_p=130\ \text{meV}$.
% For approach ({I}), we numerically integrate Eq.~\eqref{eq:int-u} and Eq.~\eqref{eq:int-xi}. 
%Sample molecules 
For approaches ({II}) and ({III}), we sample $4000$ molecular positions $\{\mathbf{r}_j\}$ with random orientations $\{\mathbf{n}_j\}$ in the molecular layer of thickness $1.0\ \mu\text{m}$ and calculate $\{g_j\}$ and $\{u_j\}$ using Eqs.~\eqref{eq:g_j_FDTD} and \eqref{eq:u_j_FDTD}.
\begin{figure} % Optional placement specifier
    \centering
    \includegraphics[width=1.0\linewidth]{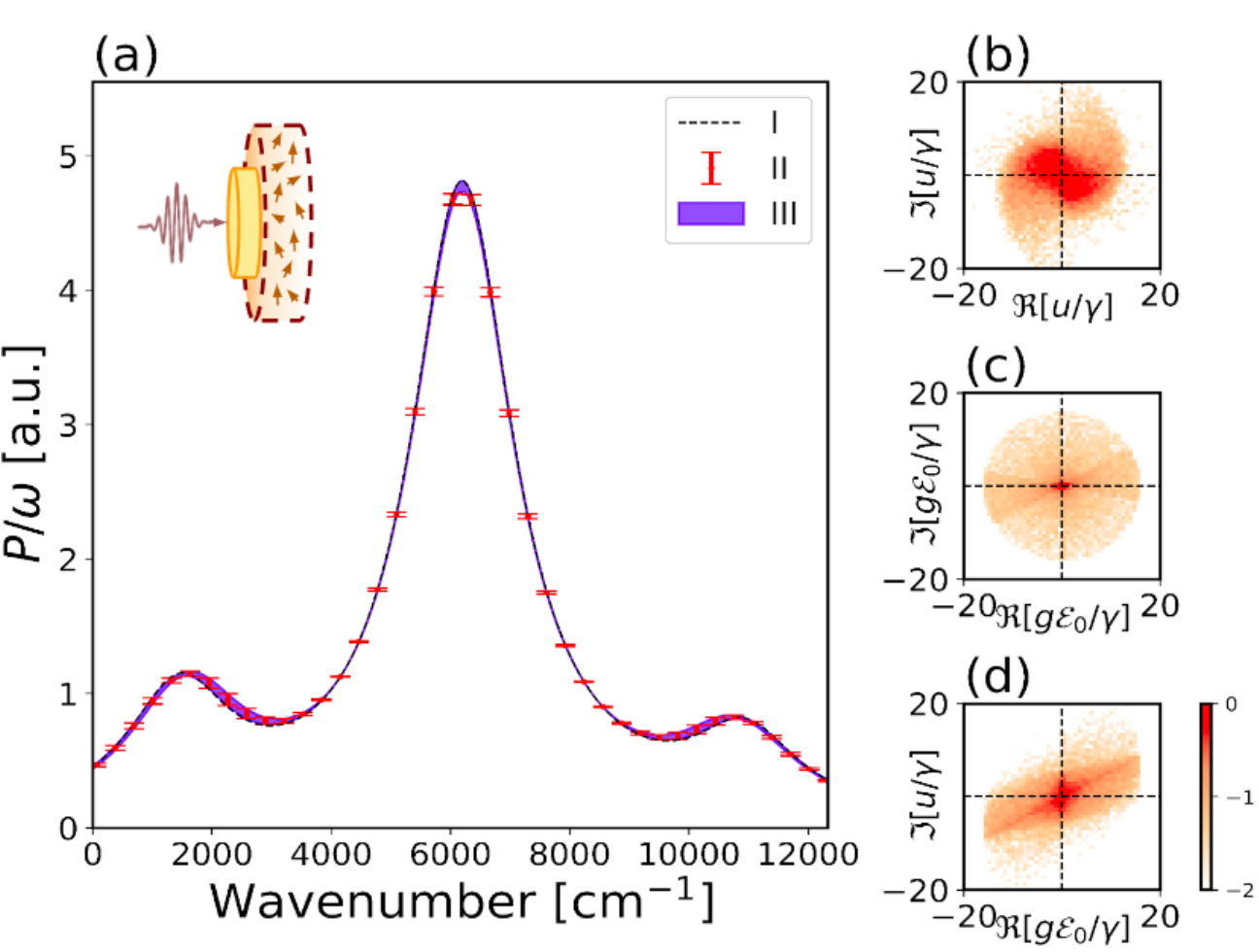} % Adjust width as needed
    \caption{
    The steady-state absorption spectrum $P(\omega)/\omega$ of molecular ensemble near a plasmonic nanodisk (as shown in the schematic figure) is calculated by ({I}) homogeneous and isotropic approximation (dashed line), ({II}) Monte Carlo method (red error bar), ({III}) the collective-mode EOM (blue shaded area).  % 4000 mole * 10 average * 10 statistics 
    Here $g_p$ is chosen to real-valued and positive and the plasmon resonance frequency is $\omega_p=6196~\text{cm}^{-1}$ and $\gamma_p = 130~\text{meV}$.
    The amplitude of the molecular transition dipole moment is chosen to be $\mu = 1.0 [\text{Debye}]$.   % $N$ = 1.73 $\times 10^{11}$
    For ({II}) and ({III}), the distribution of $\{g_j\}$ and $\{u_j\}$ based on the sampled molecules are plotted in a complex plane in (b) and (c). 
    Note that both distributions are centered at the origin, implying that $g_0=0$ and $u_0=0$.
    The correlation between $\{g_j\}$ and $\{u_j\}$ are plotted in (c), showing that the molecule-field coupling and the molecule-plasmon coupling are partially correlated with $|\xi|\approx0.5$.} \label{fig:fig3} % Optional label for cross-referencing
    % Note that the dotted line is the same spectrum with the correlation factor $\xi=0$.
    % The dashed line is classical-effective model with parameters extracted through Monte-Carlo sampling of 4000 molecules within the mask. The filled circles are brute-force Monte-Carlo simulation of the 4000 effective dipoles through exact-diagonalization. 
    % (b) cavity-molecule coupling constant $g_c$ distribution on the complex plane. (c) external-field-molecule coupling constant $g_v$ distribution on the complex plane. (d) The distribution function of $\Re{[g_v]}$ and $\Im{[g_c]}$. The $g_c/\gamma$ ( $g_v E/\gamma$) has a unit of $10^{-6}$ ($10^{-11}$). The angular frequency $\omega$ and decay rate $\gamma$ of the molecules are set to the same as $\omega_p$ and $\gamma_p$ of plasmonic cavity, where decay rate $\gamma$ = 79.8 [meV]. The dipole moment of molecule is set to 0.25 [Debye]. External field strength $E$ = 1.00 [$V/m$]. Molecule number $N$ = 1.73 $\times 10^{11}$, The molecule sampling mask is a cylinder of radius = height = 1.00 [$\mu m$] behind the plasmonic gold disk of radius = 0.25 [$\mu m$]. The color bar shows that the lighter color denotes larger distribution functions in (b-d).}
    % \wkl{[09261740]$\gamma$ = 130 [meV], $d0$ = 1.0 [Debye], $N$ = 1.73 $\times 10^{11}$} 
    % \wkl{reorder approach I II II and move b-d sub fig to the right of a}
    % \wkl{remove red dots}
    
\end{figure}

Fig.~\ref{fig:fig3}(a) demonstrates agreement across the three approaches for this system. 
% centered at zero 
Notably, the system exhibits radiative coupling disorder, as both the molecule-field coupling $\{g_j\}$ and molecule-plasmon coupling $\{u_j\}$ distributions are centered at the origin (Fig.~\ref{fig:fig3} (b,c)). 
This observation indicates that the sampled molecules follow the isotropic approximation (as assumed in Approach (I)), with $u_j$ and $g_j$ both odd under the inversion symmetry transformation $\mathbf{n} \rightarrow -\mathbf{n}$.
% correlation
In Fig.~\ref{fig:fig3}(d), the joint distribution of $\{g_j\}$ and $\{u_j\}$ clusters along a linear trend, suggesting partial correlation with $|\xi|\approx0.5$.
As a result, the simulated absorption spectrum agrees with the corresponding spectrum in Fig.~\ref{fig:fig2}(b).
Furthermore, in the absence of the incident field ($g_j=0$), the central peak will disappear, and the two side peaks recover the disorder-induced spectral splitting shown in Fig.~\ref{fig:fig1}.
\paragraph{Conclusion---}
We reveal that spectral splitting, a widely accepted signature of strong light-matter interactions, can be observed in the strong disorder regime. 
To elaborate on this observation, we develop an effective model using collective modes and demonstrate this phenomenon using electrodynamics simulations of a molecular ensemble coupled to a plasmonic nanodisk. 
Our findings represent a paradigm shift in understanding strong light-matter interactions in complex chemical environments, highlighting disorder effect as a crucial factor.
We emphasize that disorder-induced spectral splitting originates from collective dark modes, while the typical Rabi splitting arises from the bright mode.
Notably, despite their different origins, these effects appear similar in steady-state absorption spectroscopy.

Looking ahead, a key question now is how to distinguish whether a system exhibits strong coupling effects induced by polaritons or strong disorder effects dominated by the dark states. 
Beyond steady-state observations, transient dynamics offer a potential route through differentiating the relaxation and decoherence behaviors of polaritonic and dark states\cite{virgili_ultrafast_2011,perez-sanchez_simulating_2023}.
Another approach is to examine off-resonant excitation conditions to isolate distinct contributions from coherent polaritons\cite{delpo_polariton_2020}.
We will pursue these directions to harness strong light-matter interactions in diverse, disordered environments.

% \wkl{\paragraph{Feedback on the Manuscript---}

% 1. are we still calling it Rabi splitting when disorder is large, there is a inconsistency in language

% \paragraph{Effective Dipole---}
% When the coupling constants ($g_j$ and $u_j$) are identical for each group $k$ of molecules one can renormalize Eqs.~\eqref{eq:CQED-EOM-1} and \eqref{eq:CQED-EOM-2} as following:
% Suppose for each group $k$ of $M$ molecules that are indexed as $0 \leq l < M$ one has $g_j=g_{kl}=g_{k0}$ and $u_j=u_{kl}=u_{k0}$, which results $\bar{a}_j=\bar{a}_{kl}=\bar{a}_{k0}$ one can renormalize the equations with each group $k$ of molecules with a renormalized effective molecule $\tilde{\bar{a}}_k \equiv \sqrt{M} \bar{a}_{k0}$, which has complex-valued frequency $\Omega_v =\omega_v-i\gamma_v/\hbar$. The effective molecule  $\tilde{\bar{a}}_k$ has coupling constants $\tilde{g}_k \equiv \sqrt{M} g_{k0}$ and $\tilde{u}_k \equiv \sqrt{M} u_{k0}$.
% With this technique one can study the behavior of $N$ molecules with $N/M$ effective molecules.
% }
% \wkl{elaborate this and put it into supplement material}
\paragraph{Acknowledgment---}
This work was supported by the University of Notre Dame and Materials Science and Engineering fellowship. 
We are grateful to Yi-Ting Hsu for discussions.
\bibliography{reference}

\end{document}